\documentclass{article}
\usepackage{
    spconf,
    amsmath,
    graphicx,
    hyperref,
    comment,
    booktabs,
    multicol,
    multirow,
    siunitx
}
\usepackage{subcaption}
\usepackage{siunitx}
\usepackage[table]{xcolor}

\makeatletter
\let\origthebibliography\thebibliography
\let\endorigthebibliography\endthebibliography
\renewenvironment{thebibliography}[1]{%
  \origthebibliography{#1}%
  \setlength{\itemsep}{0pt}% ここを好みで調整（例: 0pt, 0.2\baselineskip など）
  \setlength{\parskip}{0pt}%
}{\endorigthebibliography}
\makeatother

\title{Sidon: Fast and Robust Open-Source Multilingual Speech Restoration for Large-scale Dataset
Cleansing}
\name{Wataru Nakata, Yuki Saito, Yota Ueda, Hiroshi Saruwatari}
\address{The University of Tokyo, Japan.}
\begin{document}
    \ninept

\setlength{\abovedisplayskip}{5pt} % 式の上部のマージン
\setlength{\belowdisplayskip}{5pt} % 式の下部のマージン
\setlength\floatsep{6pt} % 図と図の間のマージン
\setlength\intextsep{6pt} % 図と本文の間のマージン
\setlength\textfloatsep{6pt} % 図と本文の間のマージン
\setlength\abovecaptionskip{5pt} % 図とキャプションの間のマージン
\setlength{\dblfloatsep}{5pt}

    \maketitle

    \begin{abstract}
        Large-scale text-to-speech (TTS) systems are limited by the scarcity of clean,
        multilingual recordings. We introduce \textbf{Sidon}, a fast, open-source
        speech restoration model that converts noisy in-the-wild speech into
        studio-quality speech and scales to dozens of languages. Sidon consists of
        two models: w2v-BERT 2.0 finetuned feature predictor to cleanse features from noisy speech
        and vocoder trained to synthesize restored speech from the cleansed features. 
        Sidon achieves restoration performance comparable to Miipher: Google's internal speech restoration model with
        the aim of dataset cleansing for speech synthesis. Sidon is also computationally
        efficient, running up to 500\,$\times$ faster than real time on a
        single GPU. We further show that training a TTS model using a Sidon-cleansed automatic speech
        recognition corpus improves the quality of synthetic
        speech in a zero-shot setting. Code and model are released to
        facilitate reproducible dataset cleansing for the research community.
    \end{abstract}

    %-
    \vspace{-8pt}
    \section{Introduction}
    \label{sec:intro}
    \vspace{-5pt}    
    Recent advances in deep learning have demonstrated remarkable
    progress across a variety of domains, largely enabled by scaling model
    capacity and dataset size. Similar trends can be observed in speech synthesis,
    where data quantity and quality play a critical role to improve the quality
    of synthetic speech. In fact, recent text-to-speech (TTS) models are trained on tens of
    thousand hours of speech corpus~\cite{naturalspeech3, le2023voicebox} and for spoken language
    modeling, the number becomes even larger as they are trained on million
    hours of speech~\cite{whisper,Defossez2024MoshiAS}. However, unlike text or image domains, expanding
    training data by crawling in-the-wild speech samples for speech synthesis is
    particularly challenging due to the presence of noise, artifacts, and recording
    environment.

    % Beyond TTS, end-to-end spoken language models that operate directly
    % on speech are rapidly emerging~\cite{gslm,audiolm,speechgpt,seamlessm4t,gpt4o}.
    % These models perform speech understanding and generation---often with long-context
    % reasoning, translation, and dialog---without relying on intermediate text.
    % Given their dependence on large dataset, we need a method to efficiently use
    % in-the-wild speech for training these models.

    % To mitigate these issues, speech restoration models have been developed. Speech
    % restoration is a task to convert, noisy in-the-wild speech into the clean
    % speech recorded in a studio. This task often involves, speech enhancement, dereverberation,
    % super resolution, and removing codec artifacts. Notably, Miipher~\cite{miipher}
    % and Miipher-2~\cite{miipher2} have shown strong effectiveness in converting noisy,
    % real-world recordings into studio-quality speech. These models have been adopted
    % in datasets such as LibriTTS-R~\cite{librittsr} and FLEURS-R~\cite{fleursr},
    % and their benefits have been widely recognized in prior work. Nevertheless,
    % their closed-source nature restricts accessibility and limits applicability
    % to new datasets. Even though there are open-sourced models, They are often
    % monolingual, trained on limited noise conditions and small dataset. Lacks in
    % a generalization we need for apply these methods for the dataset restoration.
    To address these issues, speech restoration~\cite{liu22y_interspeech,miipher,karita2025miipher} aims to map noisy, in-the-wild speech to clean, studio-quality speech. The task typically combines speech enhancement, dereverberation, super-resolution, and codec-artifact removal. Notably, Miipher~\cite{miipher} and Miipher-2~\cite{karita2025miipher} have proven effective at converting real-world recordings into studio-quality speech, and have been adopted in datasets such as LibriTTS-R~\cite{koizumi23_interspeech} and FLEURS-R~\cite{ma24c_interspeech}. However, their closed-source nature restricts accessibility and limits applicability to new datasets. Open-source alternatives exist, but they are often monolingual and trained under limited noise conditions on relatively small datasets, which hinders generalization and reduces their utility for large-scale dataset restoration.

    In this paper, we present \textbf{Sidon}, an open-source multilingual speech
    restoration model designed for large-scale dataset cleansing. Sidon is
    computationally efficient, enabling practical use for cleansing  large corpora,
    and supports diverse languages beyond English.
    We conduct three set of experiments: (i) speech restoration on English data,
    (ii) speech restoration across 100 languages, and (iii) English TTS training
    using Sidon-processed datasets. Results show that Sidon achieves performance
    comparable to Miipher, and Sidon effectively improves the quality of
    TTS training data, thereby enhancing downstream speech synthesis quality.
    Speech samples, demo and code are available at our project page\footnote{\url{https://hf.co/spaces/Wataru/SidonSamples}}.

    \vspace{-8pt}
    \section{Related work}
    \vspace{-5pt}

    Speech restoration, also known as universal speech enhancement, aims to recover clean speech from signals degraded by noise, reverberation, bandwidth limitations, and other artifacts. VoiceFixer~\cite{liu22y_interspeech} introduced a two-stage analysis–synthesis pipeline that predicts intermediate representations from degraded speech and uses a neural vocoder to synthesize high-fidelity audio; it restores speech corrupted by noise, reverberation, clipping, and also upscales low-bandwidth signals to 44.1~kHz. Several work explored the generative modeling approach for speech restoration~\cite{selm,Guimares2025DiTSEHG,hiresldm} and demonstrated its effectiveness for improving the perceptual quality of restored speech. However, these work have focused on English, limited noise conditions, and relatively small training sets, yielding insufficient generalization for large-scale dataset cleansing.

To improve generalization, Miipher was proposed. Miipher integrates self-supervised learning (SSL) model representations from w2v-BERT~\cite{w2vbert} and linguistic conditioning from PnG-BERT~\cite{jia21_interspeech} to predict clean features that condition a neural vocoder, enabling robust restoration across diverse degradations and facilitating training high-quality TTS models from cleansed ASR datasets~\cite{miipher}. Its successor, Miipher-2, leverages the Universal Speech Model (USM)~\cite{usm} as a frozen feature extractor and trains parallel adapters~\cite{he2022towards} to predict clean USM representations from noisy inputs, improving cross-lingual generalization and scaling to million-hour corpora without explicit text or speaker conditioning~\cite{karita2025miipher}. Crucially, these systems are trained at scale (e.g., Miipher on 2{,}680 hours of English speech; Miipher-2 on 3{,}195 hours of multilingual speech), which contributes to strong generalization required for universal dataset cleansing~\cite{koizumi23_interspeech, ma24c_interspeech}. Nevertheless, both models are closed-source, and their full training data and recipes are unavailable, limiting applicability to new datasets and hindering reproducibility.

Despite these advances, most open-source methods remain English-centric, assume restricted degradation types, or are trained on comparatively small corpora, and they rarely exploit multilingual self-supervised encoders. Our work addresses these gaps by proposing Sidon: an open source speech restoration model trained on diverse noise conditions on large multilingual training dataset. 
% Sidon builds on the parametric resynthesis paradigm of Miipher and uses w2v-BERT~2.0—a 600M-parameter Conformer encoder pre-trained at scale on multilingual speech (approximately 4.5M hours spanning over 143 languages)—as the feature extractor~\cite{w2vbert2}. For training, we curate roughly 9{,}000 hours of multilingual data, enabling generalization beyond English.

    \vspace{-8pt}
    \section{Sidon}
    \vspace{-5pt}
    
    \begin{figure}
        \centering
        \includegraphics[width=0.7\linewidth]{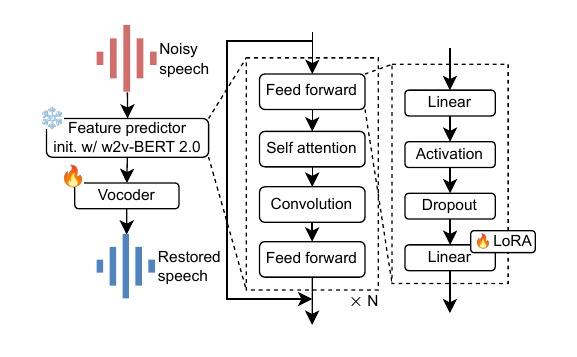}
        \vspace{-4mm}
        \caption{Overall architecture of Sidon}
        \label{fig:sidon}
    \end{figure}
    Figure~\ref{fig:sidon} shows the overall architecture of Sidon. Sidon is
    based on a parametric resynthesis framework used in previous work~\cite{miipher,karita2025miipher}.
    % Feature predictor first predicts an self-supervised learning (SSL) model feature
    % corresponding to the one extracted from clean speech.
    % Then, the predicted SSL feature is converted back to a waveform using a vocoder.

    \subsection{Feature predictor}
    \vspace{-3pt}
    Given a noisy waveform, feature predictor estimates an SSL feature extracted from
    clean speech. In order to achieve generalized speech restoration across many
    languages, it is pivotal to use an SSL model trained on large datasets from multiple
    languages. Therefore, we use w2v-BERT 2.0~\cite{Communication2023SeamlessME} as a feature predictor.
    This model is trained on 4.5M hours of speech in 143 languages, making it suitable
    for the multilingual speech restoration. Specifically, the feature predictor
    is initialized with the parameters of pretrained w2v-BERT 2.0. Then, an output
    linear layer in each feed forward network of Conformer blocks~\cite{conformer}
    are updated using low rank adaptation (LoRA)~\cite{hu2022lora} similar to \cite{karita2025miipher}.
    This strategy not only reduces training cost, but also prevents the catastrophic
    forgetting of the pretrained knowledge obtained by SSL. For the clean SSL
    model feature extraction, we extract the 8th layer hidden state from w2v-BERT 2.0.
    In previous research, it is reported that earlier layer of an SSL model contains
    acoustic features while the latter is more focused on semantic features~\cite{comparative}.
    In speech restoration, keeping the acoustic information such as speaker identity
    and prosody is also important. Therefore, we choose the 8th layer for the target
    layer.

    \vspace{-5pt}
    \subsection{Vocoder}
    \vspace{-3pt}
    The vocoder converts the predicted clean feature back to a waveform to produce
    restored speech. For this purpose, we use HiFi-GAN~\cite{Kong2020HiFi-GAN:Synthesis}
    extended with snake activation~\cite{NEURIPS2020_11604531} similar to the Descript Audio Codec
    (DAC)~\cite{kumar2023highfidelity} decoder as its model structure.

    \vspace{-5pt}
    \subsection{Dataset collection}
    \vspace{-3pt}
    \begin{table}[t]
    \centering
    \caption{List of datasets used in Sidon training. ``SF,'' ``Lang.,'' and ``Dur.' stands for sampling frequency of speech, language, and durations, respectively.}
    \scalebox{0.9}{
    \label{tab:dataset}\footnotesize
    \begin{tabular}{l|rp{2.5cm}r}\toprule
        Name & SF & Lang. & Dur. [h]  \\\midrule
        HiFi-CAPTAIN~\cite{hi-fi-captain} & 48k & ja, en & 36  \\
        EARS~\cite{richter2024ears} & 48k & en & 100 \\
        EXPRESSO~\cite{nguyen23_interspeech} & 48k & en  & 40\\
        JSUT~\cite{takamichi2020jsut} & 48k & ja & 10\\
        Bible-TTS~\cite{meyer22c_interspeech} & 48k & akan, ewe, hausa, lingala, yoruba & 80 \\
        VCTK~\cite{Yamagishi2019CSTRVC} & 48k & en & 44 \\
        JVS~\cite{takamichi2020jsut} &  24k & ja & 24 \\
        FLEURS-R~\cite{ma24c_interspeech} & 24k & 102 langs &  1.3k \\
        LibriTTS-R~\cite{koizumi23_interspeech} & 24k & en & 585 \\\midrule
        Total & & 104 & 2,219 \\\bottomrule
    \end{tabular}
    }
\end{table}
    In order to achieve speech restoration with high fidelity and good generalization,
    collecting sufficient amount of data with good recording quality is pivotal. Therefore,
    we collected multiple publicly available datasets. Table~\ref{tab:dataset}
    shows the list of the collected datasets. In Total, we collected 2,219
    hours of speech across 104 languages.

    \vspace{-5pt}
    \subsection{Training}
    \vspace{-3pt}
    We use three-stage training strategy. In the first stage, feature predictor
    is trained. In the second stage, vocoder is trained on the ground-truth
    clean SSL feature. Finally, the vocoder is finetuned on the predicted SSL
    feature obtained from the first stage.

    For the vocoder trainings in the second and third stages, only 48~kHz datasets
    were used as mixing with 24~kHz data would hurt the fidelity of the trained models
    while all the corpora listed 
    are used for training feature predictor.

    In the first stage, feature predictor is trained to minimize the mean squared
    error (MSE) loss between the predicted and target SSL features. For vocoder training
    in the second and third stages, we used a combination of MSE loss between
    generated and target mel-spectrograms, adversarial loss, and feature
    matching loss following the original HiFi-GAN paper~\cite{Kong2020HiFi-GAN:Synthesis}.

    \vspace{-5pt}
    \subsection{Differences from Miipher-2}
    \vspace{-3pt}
    Sidon follows similar architecture to Miipher-2 apart from three main differences. 
    The first difference is the SSL model. Miipher-2 uses the 13th layer hidden state of  2B parameter USM model. However, this USM model is closed source. Therefore, Sidon uses the 8th layer hidden state of open-sourced 600M parameter w2v-BERT 2.0 model. 
    The second difference is that Sidon produces 48~kHz full-fidelity speech using HiFi-GAN with snake activation as its vocoder while Miipher-2 produces 24~kHz speech using WaveFiT~\cite{wavefit}.
    Lastly, Miipher-2 was trained on 3{,}195 hours of Google internal data while Sidon is trained on 2{,}219 hours of curated high-quality public datasets.
    
    \vspace{-8pt}
    \section{Experiment}
    \vspace{-5pt}
    To validate the speech restoration performance and multilingual ability of Sidon,
    we performed following experiments: (i) speech restoration in English and multilingual settings, (ii) dataset cleansing for zero-shot TTS model training, (iii) and inference speed comparison with different batch sizes.

    \vspace{-5pt}
    \subsection{Experimental conditions}
    \vspace{-3pt}

    For clean speech dataset, we used corpora listed on Table~\ref{tab:dataset}.
    When training, valid, and test sets were specified, the split were followed.
    Otherwise, all samples were used as the training set.

    \noindent
    \textbf{Data preparation:} To train a generalized speech restoration model, diversity in degradation is important as we never know what kind of artifact present in the in-the-wild data. Therefore, we used a degradation simulation pipeline following previous work~\cite{saijo25_interspeech}.
This pipeline applies degradation in following order: reverberation,
background noise, band limitation, clipping, codec, and packet loss. Each degradation
was applied with a probability of 50\%.

    \begin{enumerate}
        \setlength{\leftskip}{-3mm}
        \setlength{\itemsep}{-1mm}

        \item Reverberation: We used pyroomacoustics~\cite{scheibler2018pyroomacoustics}
            for simulating room impulse responses (RIRs). Specifically, random RT60 and rectangular cuboid room dimensions were drawn from $\mathcal{U}(0.1,2.0)$ seconds and $\mathcal{U}(2,20)$~m respectively. Based on the drawn RT60 and room dimensions, wall absorption and maximum order of the image-source method~\cite{imagesource} were calculated using sabine's equation. Then, RIRs were simulated.

\item Background noise: We formed a noise pool from AudioSet~\cite{gemmeke2017audio},
Free Music Archive, WHAM!~\cite{Wichern2019WHAM}, FSD50K~\cite{fsd50k}, and synthetic
wind noise generated by SC-Wind-Noise-Generator~\cite{windnoise}. 
For each clean utterance, we randomly sampled a single noise recording from this pool.
The selected noise was looped to exceed the utterance duration and then truncated to match exactly, and it was added at an SNR drawn from $\mathcal{U}(-5,20)$~dB.
        % \item Background Noise: We used noises from the DNS5 challenge~\cite{}, WHAM!~\cite{Wichern2019WHAM},
        %     FSD50K~\cite{fsd50k}, the Free Music Archive\footnote{\url{https://freemusicarchive.org/home}},
        %     and wind noise simulated with the SC-Wind-Noise-Generator\footnote{\url{https://github.com/audiolabs/SC-Wind-Noise-Generator/}}~\cite{windnoise}.

        \item Band limitation: The input speech was randomly resampled at \{8,
            16, 22.05, 24, 44.1, 48\}~kHz sampling rate before being converted
            back to the original sampling rate.
        % \item Clipping: The input speech was randomly clipped by adjusting its range. The new minimum value was randomly selected between the 0th and 10th percentiles of the original signal, and the new maximum value was randomly selected between the 90th and 100th percentiles of the original signal.

        \item Clipping: The input speech was randomly clipped by setting its new
            minimum value to the value corresponding to a quantile uniformly
            chosen between the 0th and 10th percentiles, and its new maximum value
            to the value corresponding to a quantile uniformly chosen between the
            90th and 100th percentiles of the original signal.

        \item Codec: We applied the MP3 compression with a random average
            bitrate ranging from 65~kbps to 245~kbps.

        \item Packet loss: Random 9\% segments of speech were selected for packet
            loss. For each segment duration sampled from $\mathcal{U}(20,200)$~milliseconds
            were selected to be replaced with zeros to simulate packet loss.
    \end{enumerate}
    The noising pipeline was applied four times. As a result,
    we obtained roughly 9,000 hours of paired (clean, noisy) speech data.
    \begin{table}[t]
    \centering
    \caption{Evaluation results for speech restoration on LibriTTS ``test-clean'' and ``test-other'' sets. Best
    results among the restored speech are highlighted in \textbf{bold}. Values after $\pm$ indicates 95\%
    confidence intervals. Only Miipher is a text conditioned model. }
    \begin{subtable}
        [t]{\linewidth}
        \centering
        \caption{Results for ``test-clean'' set.}
        \label{tab:test_clean} \footnotesize
        \begin{tabular}{l|rrrr}
            \toprule       & WER$\downarrow$            & SpkSim$\uparrow$         & NISQA$\uparrow$                      & DNSMOS$\uparrow$                     \\
            \midrule Noisy & 0.040          & -              & 4.093 $\pm$ 0.017          & 3.179 $\pm$ 0.008          \\
            Miipher        & 0.047          & 0.942          & 4.688 $\pm$ 0.010          & 3.134 $\pm$ 0.009           \\
            Sidon (ours)          & \textbf{0.045} & \textbf{0.971} & \textbf{4.790} $\pm$ 0.010 & \textbf{3.303} $\pm$ 0.007 \\
            \bottomrule
        \end{tabular}
    \end{subtable}\hfill
    \begin{subtable}
        [t]{\linewidth}
        \centering
        \caption{Results for ``test-other'' set}
        \label{tab:test_other}\footnotesize
        \begin{tabular}{l|rrrr}
            \toprule       & WER$\downarrow$            & SpkSim$\uparrow$         & NISQA$\uparrow$                      & DNSMOS$\uparrow$                     \\
            \midrule Noisy & 0.079          & -              & 3.623 $\pm$ 0.019          & 2.949 $\pm$ 0.010           \\
            Miipher        & \textbf{0.090} & 0.930          & 4.597 $\pm$  0.011          & 3.040 $\pm$ 0.010          \\
            Sidon (ours)          & 0.095          & \textbf{0.961} & \textbf{4.698} $\pm$ 0.011 & \textbf{3.219} $\pm$ 0.008 \\
            \bottomrule
        \end{tabular}
    \end{subtable}
\end{table}
    \begin{table*}
    [t]
    \centering
    \caption{Evaluation results for speech restoration on the FLEURS test set. Best values among the restored
    speech are shown in \textbf{bold}. Due to the space constraint, we only show
    10 major languages based on Whisper's dataset distribution~\cite{whisper} but the full 100 languages results are available online at our project page.
    Please note that ``average'' indicates averaged results of ``all'' 100 languages.}
    \label{tab:placeholder}
    \scalebox{0.95}{
    \scriptsize
    \setlength{\tabcolsep}{3pt}
    \begin{tabular}{l||r|rr||r|rr||r|rr||r|rr}
        \toprule         & \multicolumn{3}{c||}{CER$\downarrow$} & \multicolumn{3}{c||}{DNSMOS$\uparrow$} & \multicolumn{3}{c||}{NISQA$\uparrow$} & \multicolumn{3}{c}{SpkSim$\uparrow$} \\
        Lang             & noisy                     & Miipher-2                      & Sidon (ours)                       & noisy                       & Miipher-2                    & Sidon (ours)                      & noisy             & Miipher-2                    & Sidon (ours)                      & noisy & Miipher-2        & Sidon (ours)          \\
        \midrule cmn     & 0.266                     & 0.287                        & \textbf{0.280}              & 3.112 $\pm$ 0.013           & \textbf{3.397} $\pm$ 0.007 & 3.374 $\pm$ 0.006          & 3.595 $\pm$ 0.026 & \textbf{4.373} $\pm$ 0.015 & 4.364 $\pm$ 0.012          & -     & 0.984          & \textbf{0.984} \\
        en               & 0.051                     & 0.061                        & \textbf{0.061}              & 2.925 $\pm$ 0.053           & 3.393 $\pm$ 0.013          & \textbf{3.466} $\pm$ 0.010 & 3.091 $\pm$ 0.104 & 4.459 $\pm$ 0.015          & \textbf{4.491} $\pm$ 0.016 & -     & \textbf{0.982} & 0.976          \\
        es               & 0.021                     & 0.022                        & \textbf{0.022}              & 3.133 $\pm$ 0.015           & 3.336 $\pm$ 0.007          & \textbf{3.399} $\pm$ 0.006 & 3.638 $\pm$ 0.030 & \textbf{4.546} $\pm$ 0.010 & 4.515 $\pm$ 0.012          & -     & \textbf{0.978} & 0.978          \\
        ru               & 0.042                     & 0.046                        & \textbf{0.043}              & 3.005 $\pm$ 0.018           & 3.307 $\pm$ 0.012          & \textbf{3.364} $\pm$ 0.009 & 3.420 $\pm$ 0.041 & \textbf{4.550} $\pm$ 0.013 & 4.490 $\pm$ 0.013          & -     & 0.984          & \textbf{0.986} \\
        fr               & 0.043                     & 0.052                        & \textbf{0.048}              & 3.125 $\pm$ 0.018           & 3.278 $\pm$ 0.014          & \textbf{3.336} $\pm$ 0.011 & 3.633 $\pm$ 0.029 & \textbf{4.428} $\pm$ 0.015 & 4.386 $\pm$ 0.013          & -     & 0.986          & \textbf{0.987} \\
        pt               & 0.028                     & \textbf{0.032}               & 0.038                       & 2.951 $\pm$ 0.022           & 3.399 $\pm$ 0.007          & \textbf{3.450} $\pm$ 0.006 & 3.364 $\pm$ 0.037 & 4.481 $\pm$ 0.010          & \textbf{4.514} $\pm$ 0.009 & -     & \textbf{0.969} & 0.964          \\
        ko               & 0.173                     & 0.183                        & \textbf{0.179}              & 2.914 $\pm$ 0.036           & 3.431 $\pm$ 0.010          & \textbf{3.465} $\pm$ 0.009 & 3.191 $\pm$ 0.030 & \textbf{4.507} $\pm$ 0.016 & 4.480 $\pm$ 0.016          & -     & 0.979          & \textbf{0.980} \\
        ja               & 0.212                     & 0.225                        & \textbf{0.213}              & 2.673 $\pm$ 0.045           & \textbf{3.479} $\pm$ 0.006 & 3.453 $\pm$ 0.006          & 2.913 $\pm$ 0.042 & \textbf{4.562} $\pm$ 0.011 & 4.472 $\pm$ 0.009          & -     & \textbf{0.959} & 0.958          \\
        tr               & 0.040                     & 0.044                        & \textbf{0.041}              & 3.069 $\pm$ 0.020           & 3.417 $\pm$ 0.008          & \textbf{3.449} $\pm$ 0.007 & 3.365 $\pm$ 0.039 & \textbf{4.556} $\pm$ 0.009 & 4.524 $\pm$ 0.009          & -     & \textbf{0.973} & 0.973          \\
        pl               & 0.030                     & 0.036                        & \textbf{0.033}              & 2.912 $\pm$ 0.023           & 3.274 $\pm$ 0.013          & \textbf{3.333} $\pm$ 0.011 & 3.544 $\pm$ 0.044 & \textbf{4.621} $\pm$ 0.009 & 4.573 $\pm$ 0.009          & -     & \textbf{0.984} & 0.983          \\
        \midrule Average & 0.084                     & 0.094                        & \textbf{0.090}              & 2.910 $\pm$ 0.003           & 3.352 $\pm$ 0.001          & \textbf{3.393} $\pm$ 0.001 & 3.252 $\pm$ 0.005 & \textbf{4.475} $\pm$ 0.002 & 4.420 $\pm$ 0.002          & -     & \textbf{0.979} & 0.979          \\
        \bottomrule
    \end{tabular}
    }
\end{table*}

    \noindent
    \textbf{DNN architecture:} For feature predictor, we set the LoRA adapter
    setting to $\alpha=16$, LoRA dropout of 0.1 and rank of $64$. For the
    vocoder, we set the upsampling rates to $\{8,5,4,3,2\}$, resulting in a 960 times
    upsampling for converting 50~Hz w2v-BERT 2.0 features into 48~kHz full fidelity speech
    and input channels to 1,536 to match the dimension of SSL feature. For other
    hyperparameters regarding the vocoder, we followed the same values as the
    previous work~\cite{kumar2023highfidelity}.

    \noindent
    \textbf{DNN optimization:} We used AdamW optimizer~\cite{loshchilov2018decoupled} ($\beta_{1}=0
    .9, \beta_{2}=0.999$) with learning rate of $1 \times 10^{-4}$ with weight decay
    of $0.01$. Learning rate exponential decay was applied when training the
    vocoder to stabilize training with $\gamma=0.9998$. The feature predictor was
    trained for 400k steps (four days) with batch size of 256, and the vocoder
    was pretrained for 140k steps (two days) and finetuned for 280k steps (four days)
    with batch size of 32. All trainings were conducted on eight NVIDIA H200 GPUs. The number of parameters were 198M (five million trainable parameters) for the feature predictor and 52.4M for the vocoder. In total, Sidon has 250M parameters.

    \vspace{-5pt}
    \subsection{Speech restoration evaluation}
    \vspace{-3pt}
    To evaluate speech restoration capability of Sidon, we compare Sidon against
    Miipher on two different settings: English and multilingual speech
    restoration.

    In the English setting, we used ``test-clean'' and ``test-other'' subsets of LibriTTS~\cite{libritts}.
    Even though we do not have direct access to the Google's internal Miipher
    model, restored samples on ``test-clean'' and ``test-other'' subsets are available as
    LibriTTS-R. We used these restored samples for evaluation.
    Please note that Miipher is a text-conditioned model while Sidon is not. Therefore, Miipher uses ground-truth transcript upon restoration.

    In the multilingual setting, we used the test set of FLEURS~\cite{fleurs}.
    Similar to LibriTTS-R, FLEURS-R contains Miipher-restored samples in the test
    set. Note that Miipher used for restoring FLEURS in our experiment differed
    from the one used in LibriTTS-R and it had identical architecture to Miipher-2~\cite{karita2025miipher}. 
    However, details of the model used in FLEURS-R such as training dataset are not disclosed in any
    papers. We refer to this model as ``Miipher-2'' in this paper for convenience.
    We noticed that FLEURS-R lacks approximately 10\% of samples from FLEURS.
    Therefore, we only used test set samples included in FLEURS-R. 
    Unlike Miipher, Miipher-2 is not a text-conditioned model. Therefore, we can expect fair comparison between Sidon and Miipher-2 regarding the text conditioning.
    Evaluation was conducted on 100 languages selected from 102 languages in FLEURS as we
    could not find any publicly available ASR models supporting Filipino (fil\_ph) and Oriya (or\_in).

    We used four metrics for evaluation. For spoken content preservation, we
    used word error rate (WER) in the English setting and character error rate (CER) in
    the multilingual setting. These metrics measure distance between transcribed results
    by an ASR model against the ground-truth scripts. For the ASR model, \texttt{mms-1B-all}~\cite{JMLR:v25:23-1318} available
    on HuggingFace\footnote{\url{https://hf.co/facebook/mms-1b-all}}, which covers
    1,162 languages with its model size of 1B parameter. For restoration quality,
    we used NISQA~\cite{nisqa} and DNSMOS~\cite{dnsmos} quality predictors.
    These are commonly used for quantifying the quality of restored speech. Specifically,
    we used overall quality scores predicted by NISQA and DNSMOS. For speaker preservation,
    we used cosine similarity of speaker embeddings (SpkSim) those extracted from input
    noisy speech and the restored speech. For speaker embedding extraction, we
    used \texttt{wavlm-base-plus-sv} available on HuggingFace\footnote{\url{https://hf.co/microsoft/wavlm-base-plus-sv}}.

    % \begin{figure}[t]
    %     \centering
    %     \includegraphics[width=0.9\linewidth]{figs/comparison.pdf}
    %     \vspace{-6pt}
    %     \caption{Mel-spectrogram comparison between noisy input, ML-Miipher-restored and Sidon-restored.
    %     The sample is taken from the FLEURS test set. Audio samples are available on our project
    %     page.}
    %     \label{fig:spec_panel}
    % \end{figure}
    
    \vspace{-5pt}
    \subsection{TTS evaluation using cleansed in-the-wild ASR dataset}
    \vspace{-3pt}
    To investigate the effectiveness of Sidon as a preprocessing tool for TTS model training, we trained a TTS model on a Sidon-cleansed ASR corpus.
    For the corpus to be cleansed, we used TED-LIUM release 3~\cite{10.1007/978-3-319-99579-3_21}, which is a 452 hours of speech dataset designed for ASR studies. It consists of TED talks recorded mostly in a hall. Therefore it contains reverberation and noise.

    For TTS model, we used F5-TTS~\cite{chen-etal-2025-f5}, which is a zero-shot TTS
    model based on flow matching~\cite{lipman2023flow}. For comparison, we
    compared Sidon with two conventional open-source speech restoration / separation models
    for cleansing TED-LIUM release 3: VoiceFixer~\cite{liu22y_interspeech} and Demucs~\cite{demucs}. VoiceFixer is an open-source speech restoration model, and Demucs is an open-source music separation model capable of separating speech from background music; both have been used for dataset cleansing~\cite{giraldo24_iberspeech,seki2025ttsopsclosedloopcorpusoptimization}.
    We compared ``Original'' (i.e., unprocessed) TED-LIUM release 3 to
    investigate the effectiveness of dataset cleansing.
    Upon training of F5-TTS, we resampled all samples to 24~kHz and performed training
    using the base model configuration of F5-TTS. The training was performed for 750k steps on eight H200 GPUs.

    For evaluation, we performed a five-scale mean opinion score (MOS) test on the sound
    quality following previous work~\cite{miipher}. The number of raters was 120
    and each raters evaluated 16 samples. To prepare test samples,
    we used the initial 40\% of each utterance as the speaker prompt and the other
    60\% were synthesized and used for evaluation.

    \vspace{-5pt}
    \subsection{Inference speed evaluation}
    \vspace{-3pt}
To efficiently cleanse large-scale in-the-wild corpora, the cleansing pipeline must be computationally lightweight.
We therefore measured the real-time factor (RTF) of Sidon on an NVIDIA H200 using batch sizes
$\{1,2,4,8\}$ in \texttt{bfloat16}. 
The measurements were taken with 16~kHz 30 seconds speech input.

    \vspace{-8pt}
    \section{Results}
    \vspace{-5pt}
    \subsection{Speech restoration evaluation}
    \vspace{-3pt}
    Table~\ref{tab:test_clean} and \ref{tab:test_other} show the result of speech
    restoration for ``test-clean'' and ``test-other'' subsets in the English setting,
    respectively.
    The results show that Sidon achieves strong performance on ``test-clean'' and better
    performance on ``test-other'' in all metrics except for WER. Please note that Miipher is a text-conditioned model while Sidon is not. Therefore, Miipher has an advantage in
    WER as it can utilize textual information. Table~\ref{tab:placeholder}
    shows the result of multilingual speech restoration.
    In terms of sound quality, a comparison of DNSMOS and NISQA scores between the noisy inputs and the Sidon outputs shows that Sidon consistently improves the sound quality in any tested language, exhibiting its success in speech restoration. 
    On average, Sidon
outperforms Miipher-2 in CER and DNSMOS, is comparable in SpkSim, and is slightly worse in NISQA.
    Therefore, Sidon's performance is comparable to Miipher-2 despite relying on the smaller SSL model.
    This may be due to the use of a more comprehensive degradation simulation pipeline with six different degradations whereas Miipher-2 only simulates reverberation, background noise and codec artifacts.
    Additionally, please note that NISQA might not be well-suited for multilingual evaluation as its trained on subjective evaluation results from English speech only.
    % Figure~\ref{fig:spec_panel}
    % provides qualitative spectrograms for a representative utterance. Sidon
    % reduces background noise and restores high-frequency detail relative to the
    % noisy input and ML-Miipher.

    \vspace{-5pt}
    \subsection{ASR dataset cleansing for TTS}
    \vspace{-3pt}
    % TTS MOS table with significance markers. Fill the macros below with your results.
\providecommand{\mosOrig}{3.254 $\pm$ 0.089}
\providecommand{\mosDemucs}{3.265 $\pm$ 0.086}
\providecommand{\mosVF}{3.771 $\pm$ 0.102}
\providecommand{\mosSidon}{\textbf{4.248} $\pm$ 0.109}
% Optional markers (edit as needed): \textsuperscript{\dag}, \textsuperscript{\ddag}
\providecommand{\sidonMarks}{\textsuperscript{\dag\ddag}}

\begin{table}[t]
    \centering
    \caption{Result of MOS test of TTS model trained on TED-LIUM Release 3 with each preprocess models. Values after with 95\% confidence intervals. Best result shown in \textbf{bold}. There are statistical significance ($p$-value $<$ 0.05) between all models except for ``Original'' and ``Demucs.''}
    \label{tab:tts_result}\small
    \setlength{\tabcolsep}{5pt}

    \begin{tabular}{l|c}
        \toprule
        Preprocess model & MOS$\uparrow$  \\
        \midrule
        Original (noisy)        & \mosOrig  \\
        Demucs~\cite{demucs}    & \mosDemucs \\
        VoiceFixer~\cite{liu22y_interspeech} & \mosVF \\
        Sidon (ours)            & \mosSidon\\
        \bottomrule
    \end{tabular}
\end{table}

    Table~\ref{tab:tts_result} shows the result of TTS model training with Sidon-cleansed dataset. The result shows that Sidon outperforms other conventional open-source restoration / separation models
    and the original noisy data in terms of the quality of synthetic speech. This indicates that Sidon can effectively
    improve the quality of training data, thereby enhancing the performance of
    downstream TTS task.

    \vspace{-5pt}
    \subsection{Inference speed evaluation result}
    \vspace{-3pt}
    \begin{table}[t]
    \newcommand{\best}[1]{\multicolumn{1}{S[table-number-font=\bfseries]}{#1}}
    \centering
    \caption{Batch scaling of elapsed time and RTF on GPU (NVIDIA
    H200). Results measured with 30 seconds 16~kHz audio input.}
    \label{tab:rtf-batch} 
    \setlength{\tabcolsep}{5pt}
    % Make S columns respect \bfseries
    \sisetup{detect-weight=true, detect-inline-weight=math}
    \begin{tabular}{ l  S[table-format=1.6, round-mode=places, round-precision=6] }
        \toprule {Batch size}  & {RTF}  \\
        \midrule   1 & 0.0022602355 \\
                   2 & 0.0020965047  \\
                   4 & 0.0020500222 \\
                   8 & \bfseries 0.0019988926 \\
        \bottomrule
    \end{tabular}
\end{table}

    Table~\ref{tab:rtf-batch} summarizes Sidon’s inference efficiency (RTF) as a function of batch size. With batch size~8 on a single GPU, Sidon runs approximately 500$\times$ faster than real time implying that restoring a 1M-hour corpus would take about 2000 hours on one GPU—substantially reducing the cost of constructing large-scale studio quality speech datasets.

    \vspace{-8pt}
    \section{Conclusion}
    \vspace{-5pt}
    In this paper, we presented Sidon, an open-source multilingual speech restoration
    model designed for large-scale dataset cleansing. Sidon is computationally
    efficient, enabling practical use in large corpus cleansing, and supports
    diverse languages beyond English. Experimental results demonstrated that Sidon
    achieves performance comparable to closed-source models like Miipher and effectively
    improves the quality of training data, thereby enhancing downstream speech
    synthesis quality. We believe that Sidon will be a valuable tool for the speech
    synthesis community, facilitating the development of high-quality, large-scale
    speech datasets.

    \textbf{Acknowledgements:} This work was supported by JST Moonshot
    JPMJMS2011, JSPS KAKENHI, Grant Number 25KJ0806, Research Grant S of the Tateisi Science and Technology
    Foundation and the AIST policy-based budget project ``R\&D on
Generative AI Foundation Models for the Physical Domain.''

\bibliographystyle{IEEEtran}
\bibliography{mybib}
\end{document}